\documentclass[
 reprint,
 amsmath,amssymb,
 aps,
]{revtex4-1}

\usepackage{graphicx}
\usepackage{dcolumn}
\usepackage{bm}
\usepackage{color}
\usepackage{comment}
\usepackage{siunitx}
\usepackage{textgreek}
\usepackage[bottom]{footmisc}

\begin{document}

\title{A tunable plasma-based energy dechirper}

\author{R.~D'Arcy$^1$}
\email{richard.darcy@desy.de}
\author{S.~Wesch$^1$}
\author{A.~Aschikhin$^{1,2}$}
\author{S.~Bohlen$^{1,2}$}
\author{C.~Behrens$^1$}
\author{M.J.~Garland$^1$}
\author{L.~Goldberg$^{1,2}$}
\author{P.~Gonzalez$^{1,2}$}
\author{A.~Knetsch$^1$}
\author{V.~Libov$^{1,2}$}
\author{A.~Martinez de la Ossa$^{1,2}$}
\author{M.~Meisel$^{1,2}$}
\author{T.J.~Mehrling$^{1,3}$}
\author{P.~Niknejadi$^1$}
\author{K.~Poder$^1$}
\author{J.-H.~R{\"o}ckemann$^{1,2}$}
\author{L.~Schaper$^1$}
\author{B.~Schmidt$^1$}
\author{S.~Schr{\"o}der$^{1,2}$}
\author{C.~Palmer$^{1,4}$}
\author{J.-P.~Schwinkendorf$^{1,2}$}
\author{B.~Sheeran$^{1,2}$}
\author{M.J.V.~Streeter$^{1,5}$}
\author{G.~Tauscher$^{1,2}$}
\author{V.~Wacker$^1$}
\author{J.~Osterhoff$^1$}
\affiliation{$^1$Deutsches Elektronen-Synchrotron DESY, Notkestra{\ss}e 85, 22607 Hamburg, Germany\\
$^2$Universit{\"a}t Hamburg, Luruper Chaussee 149, 22761 Hamburg, Germany\\
$^3$Lawrence Berkeley National Laboratory, University of California, Berkeley, California 94720, USA\\
$^4$University of Oxford, Wellington Square, Oxford, OX1 2JD, UK\\
$^5$Imperial College London, Kensington,  London SW7 2AZ, UK
}

\date{\today}

\begin{abstract}
A tunable plasma-based energy dechirper has been developed at FLASHForward to remove the correlated energy spread of a 681~MeV electron bunch. Through the interaction of the bunch with wakefields excited in plasma the projected energy spread was reduced from a FWHM of 1.31$\%$ to 0.33$\%$ without reducing the stability of the incoming beam. The experimental results for variable plasma density are in good agreement with analytic predictions and three-dimensional simulations. The proof-of-principle dechirping strength of $1.8$~GeV/mm/m significantly exceeds those demonstrated for competing state-of-the-art techniques and may be key to future plasma wakefield-based free-electron lasers and high energy physics facilities, where large intrinsic chirps need to be removed.

\end{abstract}

\maketitle

The wakefield structure in a plasma-based particle accelerator \cite{Veksler1956,Chen1985} offers distinct advantages for future free-electron laser (FEL) and high energy physics (HEP) applications \cite{Adli2013}, such as strong intrinsic focusing and high accelerating gradients \cite{Blumenfeld2007}. These, in principle, allow for the stable propagation and acceleration of an injected bunch to required energies over distances orders of magnitude shorter than those possible in conventional accelerator designs. A challenge of plasma-based concepts, however, is the development of the longitudinal phase-space of the beam, accelerated in an environment that may imprint a large linear energy-time dependency -- the so-called `chirp' -- on the beam up to the GeV/mm level. Upon exit of the plasma section this large negative remanent chirp will halt FEL gain or lead to a beam size increase limiting luminosity in HEP experiments. Ideally the chirp ought to be mitigated in order to utilize plasma wakefield acceleration techniques in future facilities.

Beam loading \cite{Katsouleas1987} -- in which the steep accelerating plasma wakefield gradient observed by the electron beam is flattened due to high bunch charges, either by shaping of the bunch \cite{Tzoufras2009} or through the injection of a second bunch \cite{Manahan2017} -- has been experimentally demonstrated to minimize production of chirps in plasma \cite{Rechatin2009}. Other concepts, such as modulated plasma densities \cite{Brinkmann2017}, have also been proposed to prevent the generation of these chirps. However, recent studies indicate that it is advantageous for a beam propagating through plasma to feature a finite correlated energy spread in order to mitigate, for example, the instability that seeds hosing \cite{Mehrling2017,Lehe2017}. This effect is analogous to Balakin-Novokhatsky-Smirnov damping \cite{Balakin1983}, where a correlated energy spread mitigates transverse instabilities in linacs and storage rings. Its utilization may be necessary in future plasma-based FEL and HEP applications to conserve the required beam characteristics in the acceleration process. Allowing the generation of chirps within plasma would therefore be beneficial, with dechirping of the beam occurring in a separate section.

Removal of energy chirps using corrugated pipes \cite{Bane2012} and dielectric-based slab structures \cite{Antipov2014} has been experimentally demonstrated. To date, these structures have been shown to remove chirps on the sub-MeV/mm level, with current theoretical estimates indicating potential for growth \cite{Mihalcea2012,Wang2006,Baturin2013}. To compensate the extreme energy chirps generated in plasma-based accelerators within distances comparable or shorter than the accelerator size, a technique capable of removing chirps far exceeding those experimentally demonstrated is required. This can be achieved by taking advantage of the large electric fields inherent to the plasma acceleration process.

One such mitigation strategy, based on the observation that a beam driving a plasma wakefield  -- a so-called `driver' -- will be subjected to a decelerating longitudinal field with a particular longitudinal dependency, is explored. By carefully matching the electron plasma density to the longitudinal beam properties it is possible to reduce, and potentially remove, an initial energy chirp of a driver beam. In this Letter the utilization of such a plasma-based energy chirp compensator at the FLASHForward experiment \cite{Aschikhin2016} is described, whereby an electron bunch produced by the gun of the FLASH water-window FEL facility \cite{Tiedtke2009,Ackermann2007}, and linearly chirped using the accelerating modules and bunch compressors in the linac, is dechirped in plasma.

The time-averaged profile of the electric field in plasma, $E_z$, leading to the reduction of the energy chirp, ultimately depends on the beam phase space distribution and the plasma profile. Here we regard a flat-top plasma with a plasma electron density of $n_p$ and electron beams with densities $n_b$ on the order of $n_p$. The FEL-quality beams have an emittance smaller or equal to the matched emittance in a homogeneous plasma channel such that the relation $k_p \sigma_x \ll 1$ for the rms width, $\sigma_x$ can be maintained during the whole interaction with the plasma, where $k_p=\omega_p/c$ is the inverse plasma skindepth, $\omega_p = \sqrt{n_p e^2/m\epsilon_0}$ the plasma frequency, $e$ the elementary charge, $m$ the electron mass, $\epsilon_0$ the vacuum permittivity, and $c$ the speed of light. 

For under-dense beams, where $n_b < n_p$, linear plasma waves are driven with an on-axis electric field, given by \cite{Keinings1987}

\begin{equation}
\begin{aligned}
\frac{E_z(\zeta)}{E_0} = ~k_p^2 \int_\infty^\zeta & d\zeta'\, \cos\left[ k_p (\zeta-\zeta') \right] \times \\ 
& \int_0^\infty dr' \, k_p r' K_0(k_p r') \frac{n_b(\zeta',r')}{n_p} \; ,
\label{eq:Ez_zeta1}
\end{aligned}
\end{equation}

\noindent where $\zeta = z -ct$ is the co-moving variable, $r'$ the radius, $E_0 = \omega_p m c /e$ the cold non-relativistic wave-breaking field, and $K_0$ the modified Bessel function of the second kind.

\begin{figure}[t]
\centering
\includegraphics[width=\columnwidth]{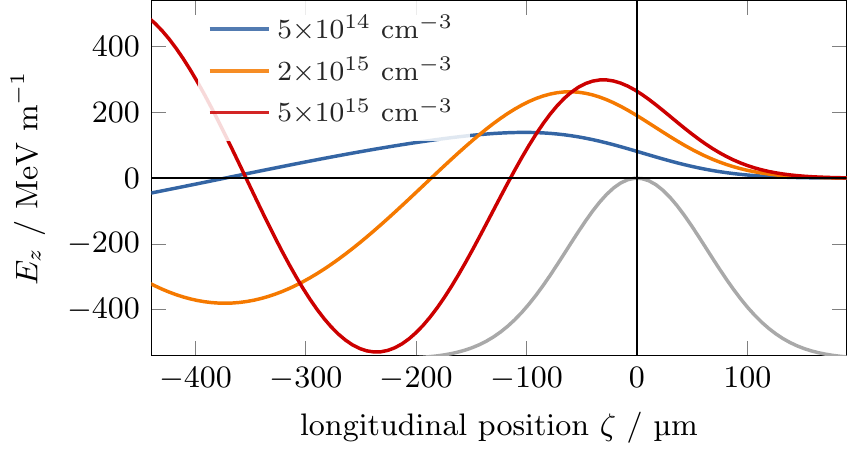}
\caption{Lineouts of the analytical electric fields, $E_z$, of three plasma wakefields (with differing electron densities) as driven by a 3D Gaussian beam. The longitudinal distribution of the electron beam is shown in grey in arbitrary units with the head of the bunch located towards the right.}
\label{fig:wakefield}
\end{figure}

For beams with a transverse Gaussian profile with $k_p \sigma_x \ll 1$,

\begin{equation}
\frac{E_z(\zeta)}{E_0} \simeq  -\log(k_p \sigma_x) \, k_p \int_\infty^\zeta d\zeta'\, \cos\left[ k_p (\zeta-\zeta') \right] \frac{2 I_b(\zeta')}{I_A} \; ,\label{eq:Ez_zeta}
\end{equation}
\\
\noindent where $I_b(\zeta')$ is the beam current profile and $I_A$ the Alfv\'en current. Such field profiles generated from a 3D Gaussian beam are illustrated in Fig.~\ref{fig:wakefield}. A characteristic beam length of $k_p l_z \lesssim 1$ ensures that the beam quasi-resonantly excites the plasma wave and that $E_z$ is monotonically increasing from the head to the tail for the majority of the beam. This electric field will, therefore, reduce the correlated energy spread of a beam with a linearly increasing energy profile from head to tail i.e.~a negative energy chirp.

For the blow-out regime, with $n_b \gg n_p$, similar considerations can be made. If, furthermore, $k_p \sigma_x \ll 1$ the majority of the beam is embedded in the generated blowout, such that $E_z$ is constant within a beam slice and, to a good approximation, only depends on $I_b$ and $n_p$ (cf.~e.g.~\cite{Lu2006,Yi2013}). While the profile of $E_z$ in the blowout regime differs from that of the quasi-linear regime, the dechirping mechanism is analogously effective.

Assuming a constant transverse rms beam size during propagation in the flat-top plasma target with length $L_p$, the energy change along the beam is given by

\begin{equation}
\Delta \gamma(\zeta) = - k_p L_p  E_z(\zeta)/E_0 \; ,
\label{eq:dechirp}
\end{equation}

\noindent where $\Delta \gamma$ denotes the change of the relativistic Lorentz factor. Hence, for a known current profile, the plasma density and target length can be experimentally tuned in order to optimize the dechirping process with the goal of minimizing the final energy chirp. Since the impact of the dechirper linearly scales with the length of the device it is convenient to characterize the dechirper strength in units of MeV/mm/m, ie. the chirp compensated over a meter-long dechirping length. These considerations determined the design of the experimental set up as described in the following methodology.

\begin{figure*}[t]
\centering
\includegraphics[width=\textwidth]{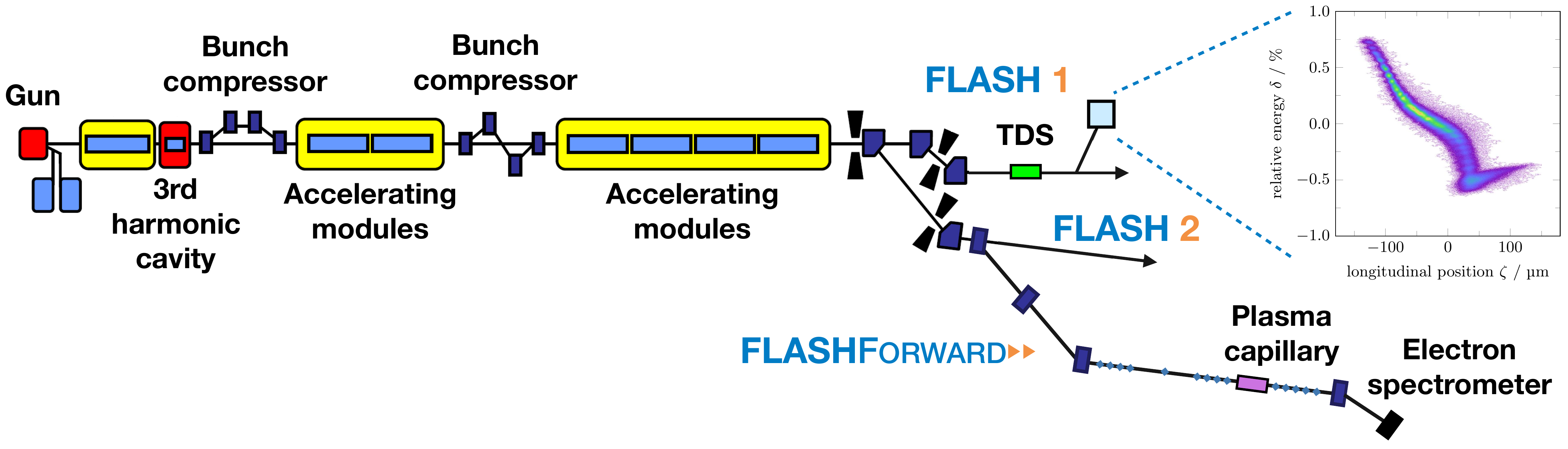}
\caption{Beamline schematic of the FLASH water-window FEL facility illustrating the RF gun and linac components used to accelerate, compress, and chirp the electron beams. The magnetic dipoles deflecting the beam into FLASHForward, as well as some components of the FLASHForward beamline itself, are also shown. The linearly chirped longitudinal phase space of the beam used in this experiment, as measured by the transverse deflection structure, is shown in the upper right inset.}
\label{fig:schematic}
\end{figure*}

The schematic in Fig.~\ref{fig:schematic} shows the layout of the FEL facility FLASH, with its 1.3 GHz superconducting accelerating structures, third harmonic cavity, magnetic bunch compressor chicanes, and S-band transverse deflection system (TDS) for longitudinal phase space characterization \cite{Loew1965,Behrens2012}. For this experiment the electron bunch was accelerated off crest in the linac to 681~MeV and compressed in the chicanes from an initial rms bunch length of 1.95~mm. The TDS was used to establish a stable and reproducible machine working point where the required negative linear energy chirp in longitudinal phase space (typical for a PWFA scheme) was met. Using the TDS a beam with an rms bunch length of \SI{63}{\micro m} and rms energy spread of 0.56$\%$ was measured. This corresponds to a chirp of 60.5~MeV/mm. The longitudinal profile of the beam can be seen in the upper right inset of Fig.~\ref{fig:schematic}.

Once longitudinal characterization of the bunch with the TDS was complete the beam was  transported to the FLASHForward beamline through a series of magnetic doglegs, with the optics set to maintain the longitudinal bunch properties. The transverse beam size of the electron beam at the interaction point (IP) was minimized using a quadrupole scan matching routine whereby optics upstream of the IP were varied, with the resulting change in beam size observed using an Yttrium Aluminium Garnet (YAG) profile screen. For the matching routine the profile screen was moved to the position of the plasma capillary by a mover system with a total of six degrees of freedom in translation and rotation. The matching routine minimized the beam to a transverse size of \SI{60}{\micro m}~$\times$~\SI{20}{\micro m}.

For the measured longitudinal and transverse beam parameters Eq.~\ref{eq:Ez_zeta1} can be used to estimate the electron plasma density required to fully dechirp the beam. The head-to-tail energy difference of the beam is approx.~7~MeV and, therefore, a decelerating field magnitude of approx.~210~MV/m at the rear of the beam is required to compensate the energy difference in a 33~mm long plasma capillary (cf.~Eq~\ref{eq:dechirp}). Figure~\ref{fig:wakefield} shows $E_z$ along the beam for three different plasma density values, calculated using Eq.~\ref{eq:Ez_zeta1} for a wakefield driven by a bunch with the measured beam parameters. According to the quasi-linear plasma excitation model a density value of 2$\times$10$^{15}$~cm$^{-3}$ approximately fulfils this condition. The peak electron bunch density for these measured beam parameters is approx.~$1.6 \times 10^{15}$~cm$^{-3}$, therefore the beam is under-dense compared to the plasma density estimated to provide maximum dechirping. As such it is reasonable to use the quasi-linear relation of Eq.~\ref{eq:Ez_zeta1} to estimate the wakefield magnitude.

\begin{figure}[t]
\centering
\includegraphics[width=\columnwidth]{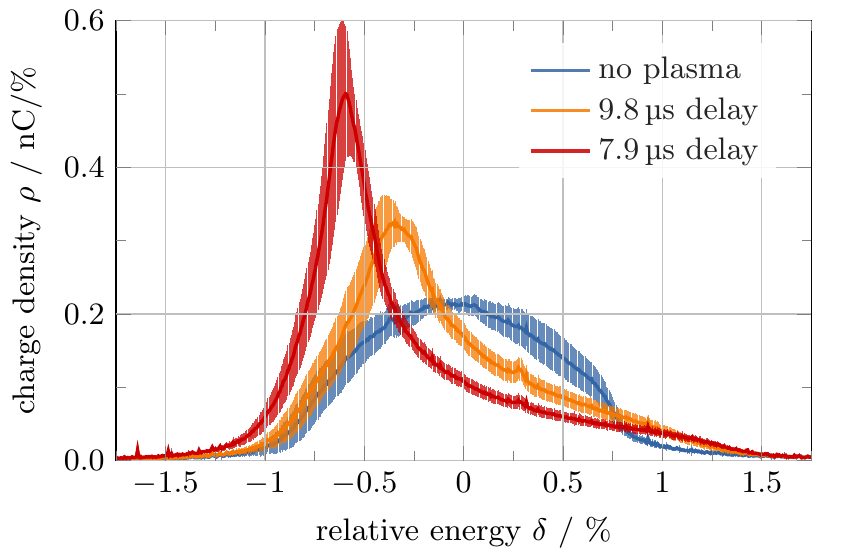}
\caption{A series of energy spectra, as recorded by the optical system surrounding the dipole spectrometer, for no interaction with plasma as well as two  dechirping plasma densities. The standard deviation for each energy slice -- an average over 50 consecutive shots -- is shown by the error ranges.}
\label{fig:spectradata}
\end{figure}

A plasma capillary, 33~mm in length and with a 1.5~mm diameter, was driven to the IP using the mover. Lossless transmission of the chirped bunch was confirmed by measuring a consistent charge of $300 \pm 2$~pC upstream and downstream of the IP. The chirped bunch was then captured by a quadrupole triplet immediately downstream of the IP and transported to a dipole spectrometer, used to disperse the chirped bunch in energy. The dispersed beam then impinged on a fluorescence screen (of Lanex Fine type) with the emitted light captured and imaged on a high resolution CCD camera with 1~Hz repetition rate. The energy spectrum of the chirped driver bunch with no plasma interaction can be seen in Fig.~\ref{fig:spectradata}, the rms of which is comparable to that measured by the TDS in FLASH1.

The chirped bunch was then injected into a plasma with a fixed average density. The plasma was generated by filling the capillary with Argon gas with a flow rate of 10~mbar~l/s and then igniting the gas to create a plasma using a 400~ns long, nearly flat-top current pulse from a 4.1~nF capacitance pulse forming network charged to 25~kV and switched by a thyratron. Once the discharge pulse ends the density of plasma electrons exponentially decays due to plasma recombination and expansion into vacuum with a lifetime on the \SI{}{\micro s}-level \cite{Loisch2018,Kallos2008}. The plasma density can, therefore, be controlled by delaying the arrival time of the electron beam relative to the discharge, with the electron beam experiencing lower densities at ever longer times after discharge. By observing a reduction of beam width on the dipole spectrometer for variable delay a maximum dechirping effect was seen at approximately \SI{8}{\micro s} after discharge, at which point the electron plasma density is optimal for dechirping. In addition to the energy distribution after no plasma interaction, Fig. \ref{fig:spectradata} shows two spectra for differing discharge delay times demonstrating the tunable plasma dechirping effect. For each delay time 50 consecutive shots were recorded in order to provide a sample size large enough to quantify the stability of the incoming beam.

\begin{figure}[t]
\centering
\includegraphics[width=\columnwidth]{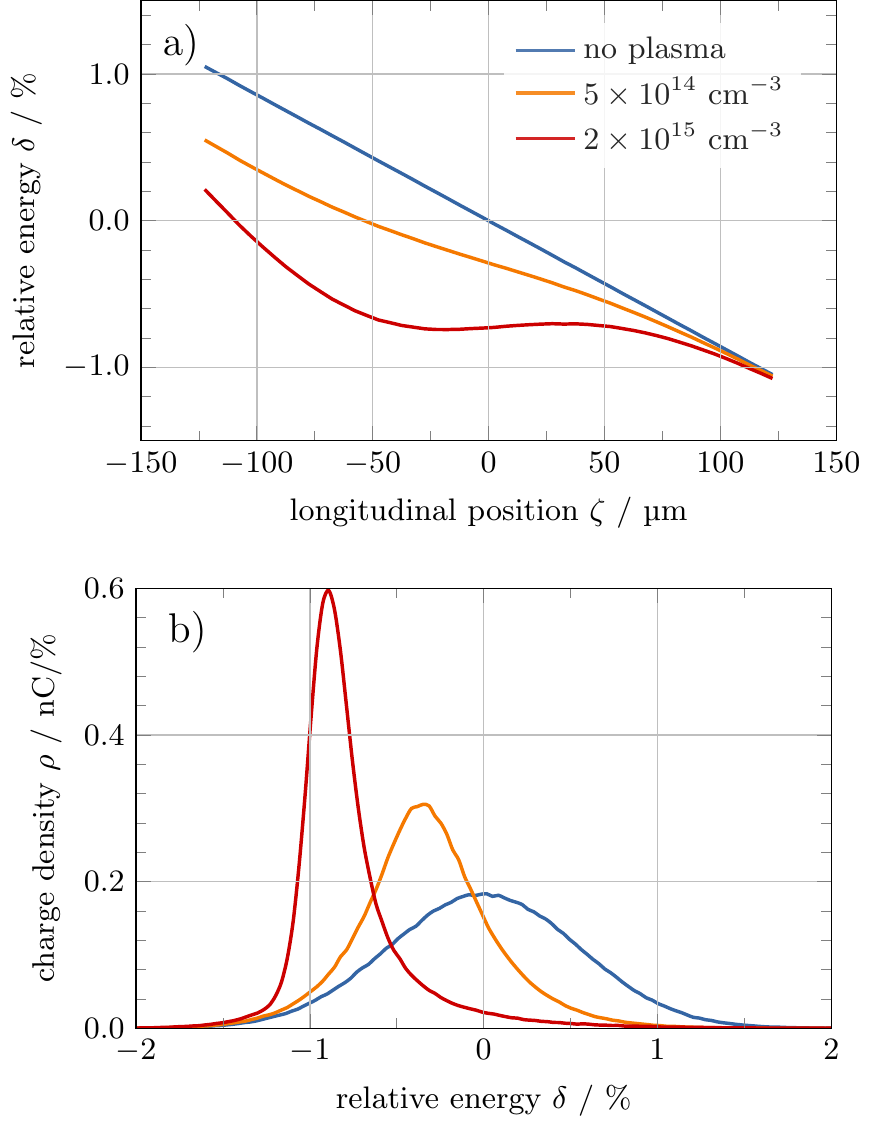}
\caption{Simulations of a) longitudinal phase space centroids, and b) collapsed energy spectra for no plasma interaction as well as two dechirping plasma densities. The three simulated spectra are equivalent to the experimental spectra shown in Fig.~\ref{fig:spectradata}.}
\label{fig:spectrasim}
\end{figure}

Using the experimentally derived electron beam parameters it is possible to calculate the expected dechirping effect for this driver for plasma densities corresponding to the delay times in Fig.~\ref{fig:spectradata}. These calculations were performed using the quasi-static three-dimensional particle-in-cell code HiPACE \cite{Mehrling2014}, whereby a 3D Gaussian representation of the bunch was propagated over a 33~mm flat-top plasma length. The plasma density was varied and the resulting longitudinal phase spaces, and therefore the energy spectra, were numerically simulated. The maximum dechirping effect in simulation was observed at a density of approx. $2 \times 10^{15}$~cm$^{-3}$, in agreement with that of Eq.~\ref{eq:Ez_zeta1}. The simulated spectrum for this maximum dechirping density, as well as that of an intermediate dechirping density, can be seen in Fig.~\ref{fig:spectrasim}. These spectra are analogous to the experimental spectra shown in Fig.~\ref{fig:spectradata} and demonstrate good agreement with data.

\begin{figure}[t]
\centering
\includegraphics[width=\columnwidth]{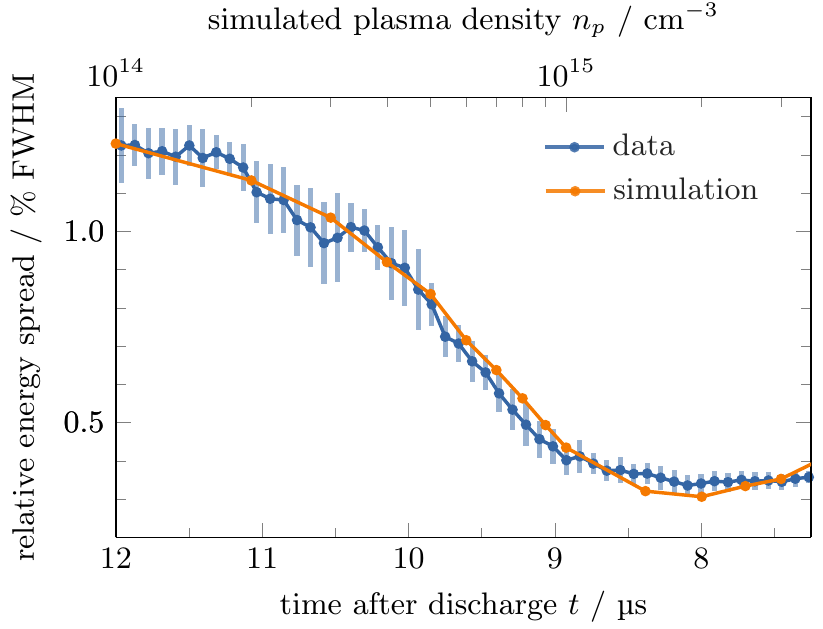}
\caption{The FWHM of the chirped bunch energy as a function of discharge time relative to the arrival time of the electron bunch. The standard deviation, representing the shot-to-shot fluctuations per delay step, is plotted. Simulated FWHM of the chirped bunch energy spectra as a function of electron plasma density over the identical range are shown for comparison.}
\label{fig:curvedata}
\end{figure}

The discharge delay time was then scanned over a wide range in steps of 92.3~ns, starting at a time before the discharge and ending when no further perturbative effect was observed on the chirped bunch. The results of this experimental scan around the delay time region of interest can be seen in Fig.~\ref{fig:curvedata}. The dechirping effect is indicated by a decrease in the FWHM of the bunch energy spectra (chosen over the rms due to the asymmetric nature of the distributions) as a function of time after discharge. The effect reaches a maximum at approx.~\SI{8}{\micro s}. The dechirping magnitude at this delay time reduces the projected energy spread of the chirped bunch from 1.31$\%$ to 0.33$\%$ over 33~mm of plasma. This reduction corresponds to a field strength of $202 \pm 18$~MeV/m at the rear of the bunch, in agreement with the magnitude of the electric field observed by the tail of the bunch as derived using the analytic formalism of Eq.~\ref{eq:Ez_zeta1} to be 210~MeV/m. At this time after discharge the plasma density is much smaller than the length-matched density of $7 \times 10^{16}$~cm$^{-3}$, i.e.~when $k_p\sigma_z = 1$, at which point maximum electric field gradients are expected. In the experimental density regime the bunch length is short compared to the plasma wavelength, resulting in the majority of the bunch experiencing the linear and monotonically increasing part of the electric field as originally suggested in Fig.~\ref{fig:wakefield}. In this case the linear chirp over the bunch centroid, previously measured as 60.5~MeV/mm, was fully compensated over a 33 mm plasma length for the 300~pC bunch, implying a dechirping strength of 1.8~GeV/mm/m. The utilization of plasma waves to compensate for chirps in our experiment enabled a dechirping strength significantly greater than previously demonstrated and has the potential to compensate even greater chirps in shorter distances in future experiments.

An additional observation from the data of Fig.~\ref{fig:curvedata} is that the standard deviation from 50 consecutive shots, indicated by the error range, decreases towards a minimum at maximum dechirping. This suggests that the implementation of the plasma dechirper in the experimental setup does not decrease the stability of the incoming beam. This stability most likely stems from the full decoherence within a few betatron periods of the beam when resonantly driving a plasma wave in the quasilinear regime \cite{Lehe2017,Martinez2018}, suppressing any hosing effects. As such it is implied that there will be no further growth of transverse instability for longer plasma capillaries.

Simulations of the chirped bunch interacting with plasma over the entire density range in Fig.~\ref{fig:curvedata} were then performed in HiPACE. The results of these simulations are displayed in Fig.~\ref{fig:curvedata}. A comparison between the profile and absolute values of dechirping for both the experimental and simulated data sets shows excellent agreement within both errors and the energy resolution of the dipole spectrometer, supporting the interpretation that the effect observed in data is indeed a plasma-induced dechirping of the bunch. The less pronounced inflexion point in the experimental data at higher plasma densities is likely due to a variable uncorrelated energy spread over the length of the bunch saturating the dechirping effect around its maximum.

In summation, a tunable plasma dechirper with a maximum dechirping strength of $1.8$~GeV/mm/m was successfully implemented in an experiment carried out at FLASHForward, DESY. By carefully selecting a plasma density at which the majority of the bunch sees a monotonically increasing $E_z$ the initial negative energy chirp of the bunch was completely removed over the centroid with a global reduction of projected energy spread from 1.31$\%$ to 0.33$\%$ FWHM. This result constitutes the first observation of its type, describing a proof-of-principle tunable plasma dechirper. If a larger integrated dechirping effect is required the technique may be scaled up by increasing the dechirping length. Furthermore, this dechirping scheme was found to not measurably affect the stability of the incoming beam. As such it may be applied to future FEL and HEP facilities where remanent chirps lead to limited functionality. In addition, this principle may be used to mitigate the large energy chirps of electron bunches generated in plasma, thus drastically improving the applicability of plasma wakefield schemes to future experiments where a negligible correlated energy spread is required.

This work was supported by the Helmholtz Virtual Institute VH-VI-503, Helmholtz ZT-0009, and Helmholtz ARD. The authors would like to thank the FLASH Directorate for their scientific and technical support. Additionally the authors would like to thank the Accelerator Test Facility (ATF), Brookhaven National Laboratory and its staff for their valuable help in the preliminary stages of experimentation. T.J.M. acknowledges the support by the Director, Office of Science, Office of High Energy Physics, of the U.S. DOE under Contract No. DE-AC02-05CH11231.

\bibliography{xampl}
\bibliographystyle{apsrev4-1}

\end{document}